\documentclass[10pt,twocolumn,letterpaper]{article}

\usepackage{iccv}

\usepackage[utf8]{inputenc} 
\usepackage[T1]{fontenc}    
\usepackage{url}            
\usepackage{booktabs}       
\usepackage{amsfonts}       
\usepackage{nicefrac}       
\usepackage{microtype}      
\usepackage{lipsum}

\usepackage{graphicx}%
\usepackage{amsmath,amssymb,amsfonts}%
\usepackage{amsthm}%
\usepackage{mathrsfs}%
\usepackage[title]{appendix}%
\usepackage{xcolor}%
\usepackage{textcomp}%
\usepackage{manyfoot}%
\usepackage{booktabs}%
\usepackage{algorithm}%
\usepackage{algorithmicx}%
\usepackage{algpseudocode}%
\usepackage{listings}%
\usepackage{mathtools}
\usepackage{array}
\newcolumntype{P}[1]{>{\centering\arraybackslash}p{#1}}
\usepackage{hyperref}       
\usepackage{cleveref}
\usepackage[T1]{fontenc}

\iccvfinalcopy
\begin{document}

\title{NBM: an Open Dataset for the Acoustic Monitoring of Nocturnal Migratory Birds in Europe}

\author{
  Louis Airale$^1$ \\
  NBM-Nocturnal Bird Migration\\
  404 chemin du casino\\
  Saint-Clément-des-Baleines, 17590 France \\
  \texttt{laerien@gmail.com} \\
   \and
  Adrien Pajot$^1$ \\
  NBM-Nocturnal Bird Migration\\
  404 chemin du casino\\
  Saint-Clément-des-Baleines, 17590 France \\
  \texttt{pajot.adrien@wanadoo.fr} \\
  \and
  Juliette Linossier \\
  Biophonia \\
  NBM-Nocturnal Bird Migration\\
  Sualello\\
  Oletta, 20232 France \\
}

\maketitle

\begin{abstract}
The persisting threats on migratory bird populations highlight the urgent need for effective monitoring techniques that could assist in their conservation.
Among these, passive acoustic monitoring is an essential tool, particularly for nocturnal migratory species that are difficult to track otherwise.
This work presents the Nocturnal Bird Migration (NBM) dataset, a collection of 13,359 annotated vocalizations from 117 species of the Western Palearctic.
The dataset includes precise time and frequency annotations, gathered by dozens of bird enthusiasts across France, enabling novel downstream acoustic analysis.
In particular, we prove the utility of this database by training an original two-stage deep object detection model tailored for the processing of audio data.
While allowing the precise localization of bird calls in spectrograms, this model shows competitive accuracy on the 45 main species of the dataset with state-of-the-art systems trained on much larger audio collections.
These results highlight the interest of fostering similar open-science initiatives to acquire costly but valuable fine-grained annotations of audio files.
All data and code are made openly available.
\end{abstract}



\section{Introduction}
\label{sec:intro}

Migratory birds have experienced a significant decline in numbers over recent decades due to various threats~\cite{both2006climate,loss2015direct,bairlein2016migratory,gilroy2016migratory,howard2020disentangling}. 
In light of the urgent need to conserve these threatened species, assessing their spatial and temporal distributions is crucial.
Currently, several methods are used to monitor bird migration, including long-term monitoring at key sites, bird ringing, citizen science initiatives or biologging~\cite{bairlein2001results,greenwood2007citizens,du2016euring,panuccio2017long,briedis2020broad,kays2022movebank,flack2022new}.
However, a significant fraction of bird species migrate at night~\cite{lowery1951quantitative,alerstam2009flight,nussbaumer2021quantifying,cooper2023songbirds}, evading these methods which focus on a limited number of diurnal individuals. 
While radars can estimate the number of individuals migrating each night in certain regions~\cite{gauthreaux2003using,gauthreaux2006monitoring,farnsworth2016characterization,van2018continental}, they cannot identify precisely the species in motion.

\footnotetext[1]{Equal contribution}
A promising solution for monitoring nocturnal migratory birds is the use of acoustic recorders that capture bird vocalizations, formally known as Passive Acoustic Monitoring~\cite{farnsworth2005flight, efford2009population}. 
This method has gained increasing popularity and has become more widespread in recent years among passionate ornithologists, particularly during the COVID-19 pandemic, facilitated by the availability of recording devices for a general audience~\cite{challeat2024dataset}.

However, analyzing several hours of recordings to detect, identify, and count migrating birds is time-consuming and requires a high level of expertise.
Automating the detection and classification by leveraging the recent and spectacular advances in computational image and sound processing is a promising avenue that has already provided compelling results, allowing for larger-scale studies~\cite{kahl2021birdnet,perch,van2024nighthawk,rauch2024birdset}.
Yet these methods are typically trained on large but only weakly annotated datasets~\cite{xenoc, macaulay}, where annotations do not indicate the precise occurrence of signals within samples that often contain many background vocalizations.
This introduces biases that can harm the classification accuracy of recognition models~\cite{michaud2023unsupervised}, or hinder their generalizability to potentially interesting use cases, such as automatic annotation of new data samples.

Recently, several finely annotated avian sound bases have also been introduced~\cite{morfi2019nips4bplus,bc_colombia,bc_amazon,bc_ne_us,bc_hawaii,bc_sierra,bc_w_us}.
These are however often limited in size or species range or only contain time annotations, completely overlooking the precious information carried by frequency.
Consequently, aside from rare exceptions such as the work of Shrestha \emph{et al}~\cite{shrestha2021bird}, bird sound recognition is consistently formulated as a \textit{multi-label classification} task on sliding audio windows.
We argue however that addressing it as an \textit{object detection} problem where vocalizations must be localized both in time and frequency (see, e.g., \Cref{fig:output_ex}) may enable unique applications such as attempting a precise counting of birds in a flock, potentially differentiated by distinct frequency signatures.
This is especially true of nocturnal migratory species, which tend to emit short, well-defined calls in a generally clear acoustic environment.
Training an object detection model on bird calls however requires a large amount of finely annotated data which are crucially missing, especially in the case of European birds which are the primary focus of the present study.
To overcome these challenges, the NBM project (for Nocturnal Bird Migration) was launched in 2020 as a crowd-sourcing initiative~\cite{binley2025quantifying}.
The objective was to share the annotation effort over privately collected recordings, complying with a standardized annotation procedure.
Nighttime calls of migratory birds were targeted, which were annotated in time and frequency, allowing for the downstream training of fine-grained object detection models.


This article introduces the result of this work, the NBM-Dataset, a collection of 13,359 meticulously annotated vocalizations, primarily capturing nocturnal flight calls and songs from 117 bird species.
To demonstrate the practical utility of this annotation approach, we develop a two-stage object detection model designed to precisely localize bird vocalizations on spectrograms.
The model is based on a FPN Faster-RCNN architecture~\cite{ren2016faster,lin2017feature}, with two key adaptations for sound event detection: we incorporate an attention mechanism to leverage the broader acoustic context of a signal, and add positional encodings to characterize the frequency and temporal scale of each sound event, as spatial invariance is not desirable in this context.

The resulting NBM detection model shows very competitive results with the BirdNet multi-label classification model~\cite{kahl2021birdnet}, despite being trained on a much smaller database.
The NBM dataset is freely available under the CC BY 4.0 license\footnote{\href{https://zenodo.org/records/14039937}{https://zenodo.org/records/14039937}}.
The code and model are also made accessible\footnote{\href{https://github.com/LouisBearing/BirdSoundClassif}{https://github.com/LouisBearing/BirdSoundClassif}}.



\section{Related Work}

General-purpose avian acoustic databases can be broadly segmented into two categories regarding the completeness of their annotated information.

Large data collections such as Xeno-Canto~\cite{xenoc} (XC) or the Macaulay library~\cite{macaulay} contain respectively around 1 and 2 million weakly-annotated audio samples, for thousands of recording hours of bird species from all over the world.
Although their labels provide no indication of the localization of the vocalizations they contain, making them unsuitable for the training of object detection systems, their large size is essential for the pre-training of many bird sound recognition models~\cite{kahl2021birdnet,perch,kahl2023overview,rauch2024birdset,swaminathan2024multi}. 

In parallel, a number of smaller-scale datasets featuring precise time and frequency annotations have been proposed in the wake of the annual BirdCLEF challenge~\cite{bc_colombia,bc_amazon,bc_ne_us,bc_hawaii,bc_sierra,bc_w_us,kahl2022overview,kahl2023overview}, that typically contain tens of thousands of labels.
As the cost of the annotation process is increased, the scope of these datasets is usually reduced in terms of species and geographical location.
The present work aims to complete this picture, with a focus on nocturnal bird calls of the Western Palearctic.
Finally, several other datasets contain localized annotations (see, for instance,~\cite{arriaga2015bird,lostanlen2018birdvox,morfi2019nips4bplus,gomez2023western,jamil2023siulmalaya}, along with the great collection of databases of~\cite{bioac_db}) but these either omit the frequency information, lack species labels, or concern different regions or species than the ones targeted in this study.

\begin{figure*}
  \centering
  \includegraphics[width=1.0\linewidth]{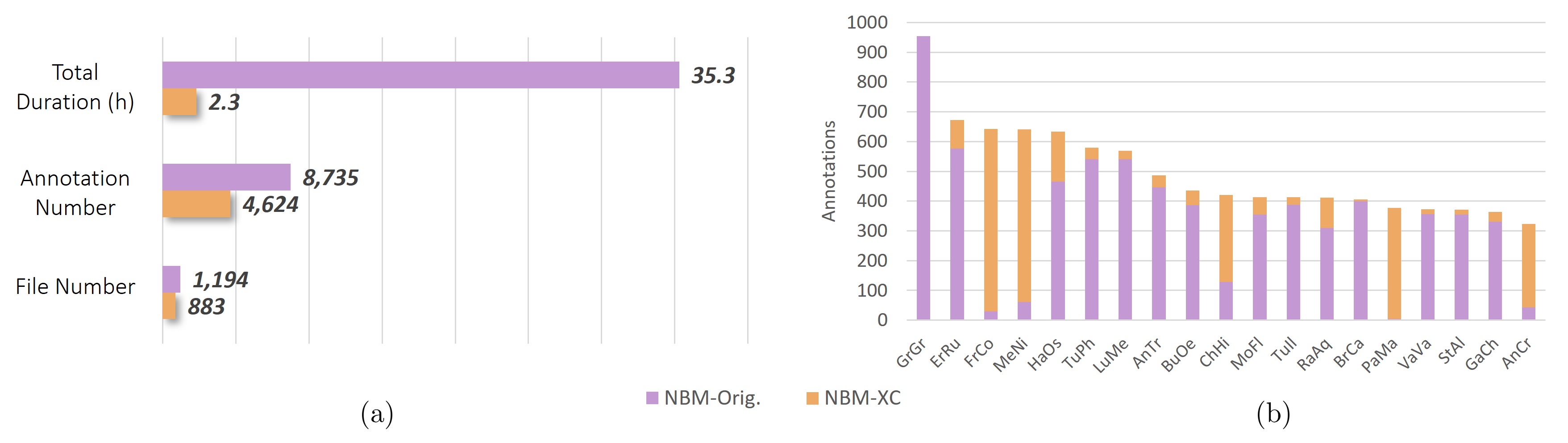}
  \caption{(a) Data volume breakdown between the original database and the XC extra base. (b) Annotation repartition between the original and XC bases for the 20 most common species. We refer the reader to \Cref{tab:sp_key_name} for the correspondence between short and proper Latin names.}
  \label{fig:ds_split}
\end{figure*}

\section{The NBM Dataset}
\label{sec:dataset}

\subsection{Data collection and Annotation}




The NBM project was launched to build a comprehensive database of vocalizations from migratory birds active at night, designed to be easily accessible for both bird enthusiasts and computer scientists. 
The initiative also aimed to encourage contributions through a simple and unified annotation process. 
A community of bird experts soon came together, sharing recordings of nighttime bird sounds along with the corresponding annotations, which provided the necessary data to develop automated systems capable of replicating the experts' annotation methodology.

This essentially implies precisely marking bird sound \textit{patterns} both in time and frequency on a spectrogram using the Audacity software~\cite{audacity}. 
These patterns were defined as the smallest identifiable segments required to recognize specific bird species, from either calls or songs. 
This method shares similarities with the 'phrases' concept used in BirdDB~\cite{arriaga2015bird}.

While there was naturally some room for personal judgment due to the diversity and complexity of bird vocalizations, as well as varying recording conditions, the specificity of nighttime soundscape—characterized by more calls than songs and fewer anthropogenic noises—helped reduce the annotation variance.

\textbf{Leveraging the French Birding Community. }
The original goal was to compile at least 100 calls from the 60 most recorded migratory bird species in Europe. 
Recognizing the heavy workload involved in annotating thousands of vocalizations, we decided to involve the French birdwatching community in a collective annotation effort. 
A call to action was issued in December 2020 and the webinar introducing the project attracted 50 participants eager to assist with the annotations. 
Following that, we developed a web platform\footnote{\href{https://nocturnal-bird-migration.com/fr}{https://nocturnal-bird-migration.com/fr}} where participants could submit audio and annotations.
The platform featured a simple system to check for consistency in annotations, and each entry was later reviewed manually.
Since the project's launch, dozens of volunteers have contributed 11,381 annotated vocalizations from 97 species.
While the majority of them are migratory birds, the collection also includes nocturnal singers, such as those in the Strigidae family, and other common species that can vocalize at night.

\textbf{Addressing the Dataset Biases. }
Involving a community of annotators to collect and annotate the data allowed to tackle a considerable dataset compilation task.
However, this approach also introduced biases and variations, particularly in terms of recording equipment and geographic locations. 
While this diversity can be a strength for improving the robustness of machine learning models given enough training data, we faced significant challenges due to an imbalanced distribution of annotations across species.
Common bird species were disproportionately represented, and some locations had an overabundance of annotations for just a few species, risking a skewed focus on a limited set of individuals in the database.
To address this, we supplemented the dataset with additional audio samples from the Xeno-Canto database~\cite{xenoc}, emphasizing the need for greater intra-class variability and better balance between species.
Since XC recordings typically include only a single class label but often contain vocalizations from multiple species, we conducted a manual time and frequency annotation process to align these samples with the standards of our database.
The resulting distribution of sample origins for the 20 most common species is shown in \Cref{fig:ds_split} (right), illustrating our endeavor to achieve a more balanced and diverse database.

\begin{figure*}
  \centering
  \includegraphics[width=1.0\linewidth]{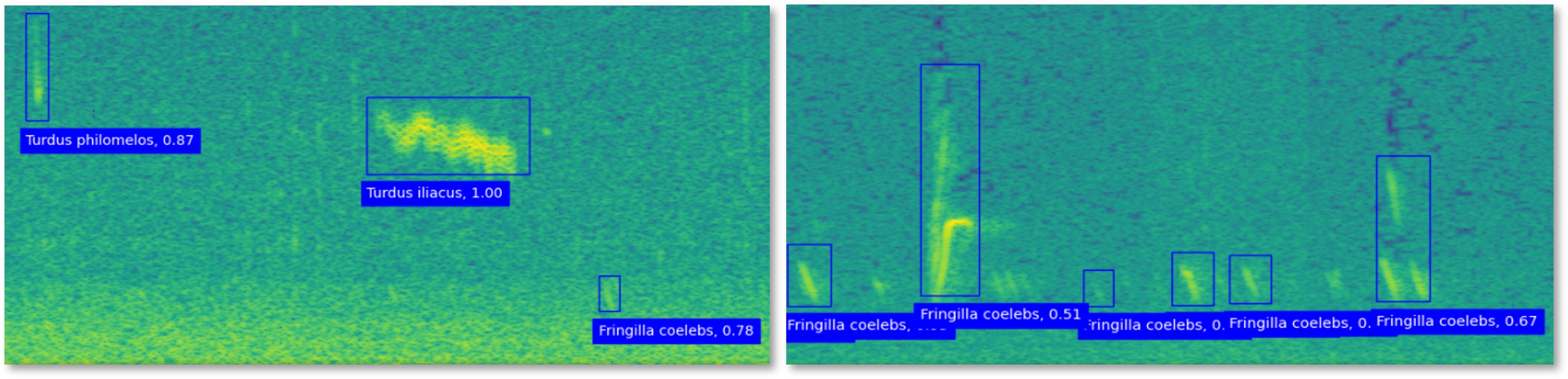}
  \caption{The object detection approach allows a fine-grained time and frequency localization of short flight calls.}
  \label{fig:output_ex}
\end{figure*}

\subsection{Database properties}

The main figures regarding the properties of the NBM database are reported in \Cref{fig:ds_split} (left), together with the share represented by its two components, the original NBM base (NBM-Orig.) and the files originating from XC (NBM-XC), in terms of recording duration, file, and annotation number.
The overall database contains almost 38 hours of recording, while total file and annotation numbers amount respectively to 2,077 and 13,359 for a total of 117 bird species.
Among these, 86 species possess at least 10 annotated calls and 56 species at least 100.
Additionally, 63 species have been annotated across at least 10 distinct files and 16 in more than 50 files.
Being an ongoing project, these figures only provide a snapshot of the NBM dataset as additional files get annotated, with the ultimate objective of covering the diversity of migratory calls of European birds as faithfully as possible.







\subsection{NBM Test Set}
Along with the training dataset described above, we introduce a test set to facilitate future research and monitor the improvements in both data collection and machine learning recognition models.
The scope of the test set is restricted to the 45 species for which we considered the amount of training data to be sufficient (see \Cref{subsec_eval_protocole} for details).
Six audio files were extracted for each species from XC and manually annotated, for a total of 270 test files.
Should the test dataset be enriched with additional samples to follow the evolution of the training set, a versioning system will be used to ensure the traceability and replicability of the results.

\section{The NBM Detection Model: an Object Detection Baseline}

The original annotation procedure of the NBM dataset allows the training of object detection models able to precisely localize bird vocalizations in an audio sample, both in time and frequency.
This approach makes the enumeration of calls possible, potentially allowing to single out individuals, thereby opening up a range of downstream applications.

This is noticeably different from the usual multi-label classification approach followed for avian sounds~\cite{kahl2021birdnet,perch,van2024nighthawk,rauch2024birdset,swaminathan2024multi}, which only predicts the presence or absence of a given species in a window of several seconds of audio.
Examples of our model's outputs on the NBM test set can be found in \Cref{fig:output_ex}, displaying the time and frequency localization of detected calls.

In this section, we present the two-stage model that we trained on the NBM dataset and which performs object detection on spectrograms.
Following the description of the model architecture, we present the results of our quantitative evaluations, including a comparison against BirdNet~\cite{kahl2021birdnet} on a multi-label classification task using sliding windows.
Our model's highly competitive results across most species highlight the effectiveness of our framework and underline the value of the NBM dataset in advancing automatic bird sound recognition.

\subsection{Model Architecture}

\begin{figure*}
  \centering
  \includegraphics[width=0.9\linewidth]{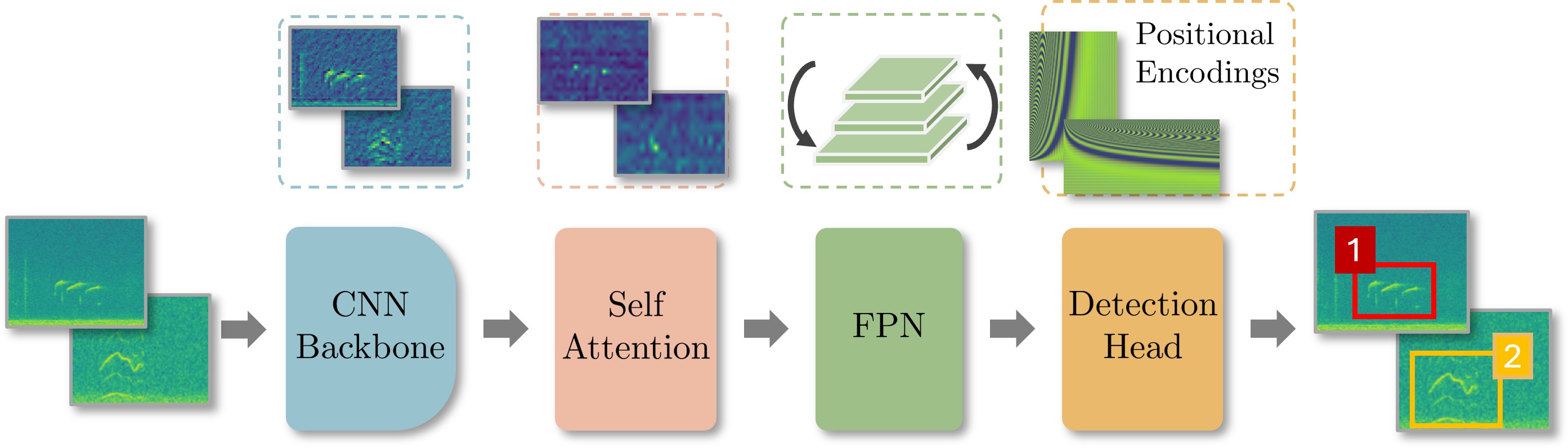}
  \caption{Overview of the presented NBM bird sound detection modular architecture.}
  \label{fig:model_archi}
\end{figure*}

Object detection neural networks can come either as one-stage~\cite{redmon2016you,lin2017focal,tan2020efficientdet,carion2020end,reis2023real} or two-stage models~\cite{girshick2015region,girshick2015fast,ren2016faster,zhu2020deformable}.
The latter involves a pass through a first network which is tasked with retrieving all potential Regions of Interest (RoIs) in an input image, corresponding to object positions.
In a second stage, another network performs classification on the RoIs.
One-stage object detectors, on the other hand, directly predict a class for each position, or anchor, in the image, including an extra background class.
Both approaches rely on an initial feature extraction through a backbone network, whose depth and expressive power directly impact the quality of the detection.

Owing to preliminary experiments, \textit{we find two-stage models to be significantly easier to train} on our avian sound detection dataset.
We hypothesize that this is due to the overrepresentation of the background class within samples, and that this class is not always correlated with an absence of signal: all sorts of background noises, starting with raindrops, will appear as peaks of intensity in the input spectrograms.
It is therefore an easier task for the model to tackle avian signal detection and classification as two separate problems.

In light of these findings, we build our detection model on the FPN-FasterRCNN (hereafter FRCNN) architecture~\cite{ren2016faster,lin2017feature}, largely modernized and tailored for the detection of bird calls.
The architecture consists of the following blocks, and is represented in~\Cref{fig:model_archi}.

\textbf{Backbone \& FPN.} 
The proposed NBM detection model is built upon a modular architecture where each of the blocks represented in~\Cref{fig:model_archi} operates independently from the others, allowing the exploration of diverse architectures within each block.
In the following, an EfficientNetV2 backbone~\cite{tan2021efficientnetv2} is used in place of the VGG16 network of the seminal FRCNN paper.
Similarly, one may explore using several variants of the Feature Pyramid Network (FPN,~\cite{lin2017feature}), whose primary goal is to feed lower layers of the network with semantic information through a top-down path~\cite{liu2018path,tan2020efficientdet}.
We experimentally find that increasing the number of output channels of a single FPN layer yields the best trade-off between accuracy and model complexity.

\textbf{Self-Attention Module.} Contrary to natural images, there is a strong correlation between bird vocalizations that can be found in a given time frame.
Equipping the model with an attention mechanism allows each local prediction to leverage the contextual information in which the call was emitted, leading to overall better predictions.
This is done by performing self-attention~\cite{bahdanau2014neural,vaswani2017attention} separately at each intermediate output from the backbone, prior to the FPN (second block of~\Cref{fig:model_archi}).

\textbf{Detection Head.} The detection head is composed of two networks: a Region Proposal Network (RPN) and a Regional CNN (RCNN), which processes and classifies the outputs from the RPN (we refer the reader to the original paper for more details).
Contrary to other vision tasks, object detection on spectrograms \textit{is not invariant on the size of the object and its position on the frequency axis}: similar patterns emitted at two different fundamental frequencies are likely to belong to two distinct species.
To account for this effect, sinusoidal positional encodings are added to the extracted RoIs (fourth block in~\Cref{fig:model_archi}).
We use an absolute positional encoding for the frequency axis and a relative encoding for the time axis.

\begin{figure*}
  \centering
  \includegraphics[width=1.0\linewidth]{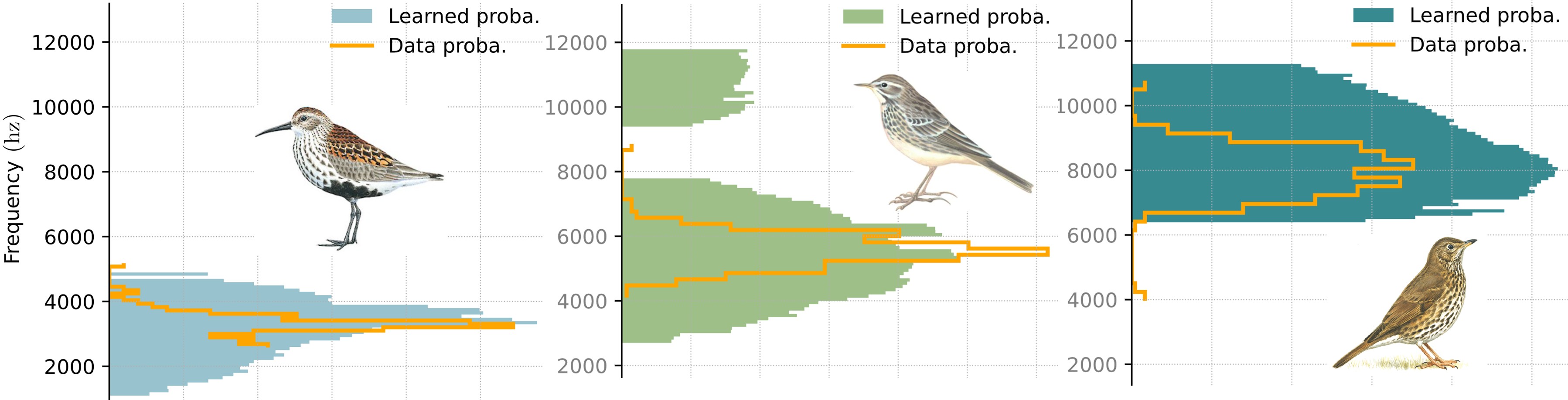}
  \caption{Learned posterior probabilities for call frequencies vs data distribution on three species: dunlin (left), tree pipit (middle), and song thrush (right). Note the alignment between ground truth and predicted frequencies, learned solely through the use of the positional encoding.}
  \label{fig:cond_probas}
\end{figure*}

The effectiveness of this strategy is illustrated in~\Cref{fig:cond_probas}.
Here we investigate how the output probability attributed to an input call varies when different frequency encodings are sampled, which mimics the extraction of the same RoI over different locations on the frequency axis.
By doing this one obtains output probabilities $p(c|f)$ corresponding to all frequency values $f$, if $c$ is the class of the input call.
Bayes' rule then provides the posterior frequency probability $p(f|c)$, learned by the model only through the use of the positional encoding:
\begin{equation*}
    p(f|c)=\frac{p(c|f)p(f)}{p(c)},
\end{equation*}
where $p(c)$ is obtained by marginalizing over $f$, all $p(f)$ taking the same value.
We report in~\Cref{fig:cond_probas} posterior probabilities averaged over all test samples for three distinct bird species.
The first observation is that these probabilities are not uniform, which shows that positional encodings indeed bias the model's output.
More importantly, the posterior probability peaks are aligned with those of the distributions measured from the training set, suggesting that the model successfully learns to exploit the information contained in the encoding to improve its predictions.


\begin{table*}[t]
\caption{Multi-label classification average precision on the considered scope of 45 species for the NBM and BirdNet models.}
\label{tab:compa_per_sp}
\resizebox{1.0\textwidth}{!}{
\begin{tabular}{l c c c c c c c c c c c c c c c}
\toprule
\textbf{Model} & GrGr & ErRu & FrCo & MeNi & HaOs & TuPh & AnTr & BuOe & ChHi & MoFl & TuIl & RaAq & PaMa & VaVa & StAl \\
\midrule
\textbf{NBM} & 0.76 & 0.33 & 0.54 & 0.90 & 0.89 & 0.93 & 0.84 & 0.76 & 1.00 & 0.97 & 0.95 & 0.72 & 0.57 & 0.69 & 0.61  \\
\textbf{BirdNet} & 0.80 & 0.42 & 0.58 & 0.99 & 0.89 & 0.52 & 0.84 & 0.86 & 0.98 & 0.87 & 0.84 & 0.44 & 0.60 & 0.79 & 0.82  \\
\midrule
GaCh & AnCr & AcHy & FiHy & EmHo & PlSq & CoCo & AtNo & FuAt & AlAr & ClGl & CyCa & EmCi & TrTo & NuAr & MuSt \\
\midrule
0.37 & 0.99 & 0.88 & 0.57 & 0.70 & 0.83 & 1.00 & 0.38 & 0.64 & 0.99 & 0.87 & 0.65 & 0.31 & 1.00 & 0.58 & 0.58 \\
0.70 & 0.95 & 0.95 & 0.24 & 0.58 & 0.95 & 1.00 & 0.75 & 0.69 & 0.80 & 0.70 & 0.96 & 0.80 & 0.86 & 0.90 & 0.61 \\
\midrule
TyAl & CaAl & TrOc & ArCi & TuMe & ChDu & TaRu & PiPi & PlAp & NuPh & TuTo & AnPr & IxMi & NyNy & & \textbf{mAP}\\
\midrule
0.97 & 0.69 & 0.78 & 0.90 & 0.3 & 0.79 & 0.37 & 0.67 & 0.26 & 1.00 & 0.69 & 0.60 & 0.73 & 0.59 & &\textbf{0.71}\\
1.00 & 0.99 & 0.99 & 0.98 & 0.51 & 0.80 & 0.66 & 1.00 & 0.71 & 0.88 & 0.73 & 0.69 & 0.61 & 1.00 & &\textbf{0.78}\\

\bottomrule
\end{tabular}
}
\end{table*}

\begin{table*}[t]
\caption{Table of correspondence between species codes and Latin names.}
\label{tab:sp_key_name}
\resizebox{1.0\textwidth}{!}{
\begin{tabular}{| c | c | c | c | c | c | c | c | c | c |}
\toprule
GrGr & Grus grus & ErRu & Erithacus rubecula & FrCo & Fringilla coelebs & MeNi & Melanitta nigra & HaOs & Haematopus ostralegus \\
TuPh & Turdus philomelos & AnTr & Anthus trivialis & BuOe & Burhinus oedicnemus & ChHi & Charadrius hiaticula & MoFl & Motacilla flava \\
TuIl & Turdus iliacus & RaAq & Rallus aquaticus & PaMa & Parus major & VaVa & Vanellus vanellus & StAl & Strix aluco \\
GaCh & Gallinula chloropus & AnCr & Anas crecca & AcHy & Actitis hypoleucos & FiHy & Ficedula hypoleuca & EmHo & Emberiza hortulana \\
PlSq & Pluvialis squatarola & CoCo & Coturnix coturnix & AtNo & Athene noctua & FuAt & Fulica atra & AlAr & Alauda arvensis \\
ClGl & Clamator glandarius & CyCa & Cyanistes caeruleus & EmCi & Emberiza citrinella & TrTo & Tringa totanus & NuAr & Numenius arquata \\
MuSt & Muscicapa striata & TyAl & Tyto alba & CaAl & Calidris alpina & TrOc & Tringa ochropus & ArCi & Ardea cinerea \\
TuMe & Turdus merula & ChDu & Charadrius dubius & TaRu & Tachybaptus ruficollis & PiPi & Pica pica & PlAp & Pluvialis apricaria \\
NuPh & Numenius phaeopus & TuTo & Turdus torquatus & AnPr & Anthus pratensis & IxMi & Ixobrychus minutus & NyNy & Nycticorax nycticorax \\

\bottomrule
\end{tabular}
}
\end{table*}

\subsection{Model Evaluation}
\label{subsec_eval_protocole}

We experimentally evaluate the performances of our detection model trained on the NBM dataset.
Beyond the validation of the architectural choices presented above, these experiments aim to show the relevance of the proposed dataset for the task of precisely localizing migratory bird calls in acoustic recordings.

\textbf{Evaluation Scope.}
As described in~\Cref{sec:dataset}, collecting the NBM dataset is an ongoing project, where the annotation effort is partly driven by the frequently re-evaluated needs for species suffering from a lack of data samples or diversity.
For a fair evaluation, it is therefore necessary to restrict the scope of the study to species that present both a sufficient number of data samples and that appear in a sufficient number of input files.
This avoids the biases caused by the over-representation of a small number of individuals in a few long training files.
We set the minimal thresholds to 100 training samples and 20 source files, to which we add three migratory species that closely approach the line (\textit{Ixobrychus minutus} \& \textit{Anthus pratensis} with 96 samples, and \textit{Nycticorax nycticorax} with 18 files and 85 samples) for a total of 45 species, which can be found in~\Cref{tab:sp_key_name}.

\textbf{Technical Details.}
All recordings were resampled to 44100~Hz, and subsequently processed through Short-Time Fourier Transform using 1024 points and a hop size of 132.
The detection model's input spectrogram resolution was then set to $375 \times 1024$, corresponding to a frequency range between 500~Hz and 13000~Hz and a duration of 3.07~s.
Such a large input size is susceptible to provide a richer acoustic context, which the model is able to exploit through its self-attention layer.
As we consider objects of smaller size than for the typical detection task, we start extracting activation maps from the second intermediate layer $\mathcal{P}_2$ of the backbone (corresponding to a reduction by a factor 2 of the input spatial dimensions), up until $\mathcal{P}_6$ (factor $2^5$, respectively).
Finally, we experimentally found that, compared to the original FRCNN configuration, additional anchor aspect ratios (akin to~\cite{lin2017focal}) and larger RPN/RCNN batch sizes provided further improvements.

\subsubsection{Performances on the Detection Task}

We calculate the mean average precision (mAP) over the 45 target species and the 271 hand-annotated XC recordings that compose the NBM test set.
A \textbf{mAP value @$\text{IoU}_{0.5}$ of 0.67} was found, which will constitute the baseline to track future improvements in model training and dataset completion.

\subsubsection{Comparison with BirdNet}

In this section, the evaluation is carried out on a multi-label classification formulation of the problem, which allows to compare the performances of the NBM model with the state-of-the-art BirdNet model.
BirdNet builds on an EfficientNet~\cite{tan2020efficientnetrethinkingmodelscaling} backbone, and is trained on a dataset that includes the whole XC database to predict the presence of bird species within 3-second time frames.
As our test set originates from XC itself, a degree of overlapping with BirdNet's training base is possible, and therefore the results presented here should be treated with caution.
However they still provide valuable insights. 
The two models are not directly comparable as their scope differs, the NBM model tackling bird sound recognition through the object detection paradigm. 
It is possible however to cast its outputs in the multi-label classification setting of BirdNet by aggregating all detected sound events in a given window and retaining the maximum detection confidence for each species.
The resulting AP scores per species and overall mAP are reported in~\Cref{tab:compa_per_sp}.
Despite the orders of magnitude of difference in the scale of their respective training sets, the NBM model results are highly competitive with those of BirdNet with 0.07 points of difference in mAP.
These results highlight the potential of a precisely annotated, medium-scale dataset (the average number of individual sound events for these 45 species is 306) compared to a much larger-scale, weakly annotated one.
This suggests that limited, well-thought-out annotation efforts can have an important and positive impact on the model's performance.

\subsection{Discussion \& Limitations}

The results from \Cref{tab:compa_per_sp} display a high AP variance among species, which mainly illustrates the sensitivity of our model to out-of-distribution test samples, e.g. in the case of \textit{Athene noctua}, \textit{Emberiza citrinella}, \textit{Turdus merula} or \textit{Pluvialis apricaria}.
For these species we observe a detrimental covariate shift, arising as a consequence of a mismatch between training and test distributions.
Here it may be an indication of a lack of diversity in the recording conditions.
Such observations are crucial to better drive the future annotation effort towards a higher intra-class diversity of training samples.
A second avenue of improvement concerns the species coverage which, when insufficient, can be a source of false positive detections as the model assigns even unknown vocalizations to the most likely class seen during training.
To mitigate this, we aim to increase the current scope of species to meet the threshold described in \Cref{subsec_eval_protocole} for all nocturnal migratory birds of Europe, along with common species that are susceptible to vocalizing during nighttime.

\section{Conclusion}

In this work, we present the NBM database, a crowd-sourced dataset of nocturnal migratory bird calls and nocturnal bird songs gathered by a community of volunteers across France and completed by hand annotations from Xeno-Canto.
This dataset contains 13,359 precise time and frequency annotations across 2,077 files, making it the first of its kind for birds of the Western Palearctic.
We demonstrate the relevance of this annotation method by training an object detector model to precisely localize avian sounds on spectrograms.
Notably, this model, trained on a much smaller database, proves to be competitive against the renowned BirdNet model in multi-label classification for the 45 most abundant species.
Although margins of improvement remain regarding the coverage of the NBM dataset, we aim to address these through an ongoing collection and annotation effort.
Given the continuous need for accurately annotated avian sound datasets to enhance automated recognition systems, we believe this dataset will be a valuable resource for the bioacoustics community.


\section*{Acknowledgements}
The authors want to give credit to all the ornithologists who made this project possible. Special thanks to the original contributors: Adrien Pajot, Aymeric Mousseau, Christophe Mercier, Frédéric Cazaban, Gaëtan Mineau, Ghislain Riou, Guillaume Bigayon, Hervé Renaudineau, Kévin Leveque, Lionel Manceau, Mathurin Aubry, Maxence Pajot, Nidal Issa, Willy Raitière, and equal thanks to all anonymous further contributors.
The authors also want to thank Natural Solutions developers and other coding volunteers and particularly: Rhandy Grard, Gaëtan Duhamel, Javier Blanco, Ludovic Descateaux, Hervé Aymes.
The authors also want to thank the members of the NBM association, Hervé Renaudineau and Léo Papet, and Paul Peyret for his reading and advice.
Additionally, the authors want to highlight the support from the French biacoustics experts Yves Bas and Maxime Cauchoix.
And finally, because of the complexity of crowd-sourcing projects, the authors want to warmly thank all the contributors who provided even punctual support through the project's communication channels.

\subsection*{Data Availability}

All data mentioned in this work, and which support the results presented here are freely accessible at the following address: \href{https://zenodo.org/records/14039937}{https://zenodo.org/records/14039937}.

\bibliographystyle{unsrt}
\bibliography{main}

\begin{thebibliography}{10}

\bibitem{both2006climate}
Christiaan Both, Sandra Bouwhuis, CM~Lessells, and Marcel~E Visser.
\newblock Climate change and population declines in a long-distance migratory bird.
\newblock {\em Nature}, 441(7089):81--83, 2006.

\bibitem{loss2015direct}
Scott~R Loss, Tom Will, and Peter~P Marra.
\newblock Direct mortality of birds from anthropogenic causes.
\newblock {\em Annual Review of Ecology, Evolution, and Systematics}, 46(1):99--120, 2015.

\bibitem{bairlein2016migratory}
Franz Bairlein.
\newblock Migratory birds under threat.
\newblock {\em Science}, 354(6312):547--548, 2016.

\bibitem{gilroy2016migratory}
James~J Gilroy, Jennifer~A Gill, Stuart~HM Butchart, Victoria~R Jones, and Aldina~MA Franco.
\newblock Migratory diversity predicts population declines in birds.
\newblock {\em Ecology letters}, 19(3):308--317, 2016.

\bibitem{howard2020disentangling}
Christine Howard, Philip~A Stephens, James~W Pearce-Higgins, Richard~D Gregory, Stuart~HM Butchart, and Stephen~G Willis.
\newblock Disentangling the relative roles of climate and land cover change in driving the long-term population trends of european migratory birds.
\newblock {\em Diversity and Distributions}, 26(11):1442--1455, 2020.

\bibitem{bairlein2001results}
F~Bairlein.
\newblock Results of bird ringing in the study of migration routes and behaviour.
\newblock {\em Ardea}, 89(1):7--19, 2001.

\bibitem{greenwood2007citizens}
Jeremy~JD Greenwood.
\newblock Citizens, science and bird conservation.
\newblock {\em Journal of Ornithology}, 148(Suppl 1):77--124, 2007.

\bibitem{du2016euring}
Christopher~R Du~Feu, Jacquie~A Clark, Michael Schaub, Wolfgang Fiedler, and Stephen~R Baillie.
\newblock The euring data bank--a critical tool for continental-scale studies of marked birds.
\newblock {\em Ringing \& Migration}, 31(1):1--18, 2016.

\bibitem{panuccio2017long}
Michele Panuccio, Beatriz Mart{\'\i}n, Michelangelo Morganti, Alejandro Onrubia, and Miguel Ferrer.
\newblock Long-term changes in autumn migration dates at the strait of gibraltar reflect population trends of soaring birds.
\newblock {\em Ibis}, 159(1):55--65, 2017.

\bibitem{briedis2020broad}
Martins Briedis, Silke Bauer, Peter Adam{\'\i}k, Jos{\'e}~A Alves, Joana~S Costa, Tamara Emmenegger, Lars Gustafsson, Jaroslav Kole{\v{c}}ek, Milo{\v{s}} Krist, Felix Liechti, et~al.
\newblock Broad-scale patterns of the afro-palaearctic landbird migration.
\newblock {\em Global Ecology and Biogeography}, 29(4):722--735, 2020.

\bibitem{kays2022movebank}
Roland Kays, Sarah~C Davidson, Matthias Berger, Gil Bohrer, Wolfgang Fiedler, Andrea Flack, Julian Hirt, Clemens Hahn, Dominik Gauggel, Benedict Russell, et~al.
\newblock The movebank system for studying global animal movement and demography.
\newblock {\em Methods in Ecology and Evolution}, 13(2):419--431, 2022.

\bibitem{flack2022new}
Andrea Flack, Ellen~O Aikens, Andrea K{\"o}lzsch, Elham Nourani, Katherine~RS Snell, Wolfgang Fiedler, Nils Linek, Hans-G{\"u}nther Bauer, Kasper Thorup, Jesko Partecke, et~al.
\newblock New frontiers in bird migration research.
\newblock {\em Current Biology}, 32(20):R1187--R1199, 2022.

\bibitem{lowery1951quantitative}
George~H Lowery.
\newblock A quantitative study of the nocturnal migration of birds.
\newblock 1951.

\bibitem{alerstam2009flight}
Thomas Alerstam.
\newblock Flight by night or day? optimal daily timing of bird migration.
\newblock {\em Journal of Theoretical Biology}, 258(4):530--536, 2009.

\bibitem{nussbaumer2021quantifying}
Rapha{\"e}l Nussbaumer, Silke Bauer, Lionel Benoit, Gr{\'e}goire Mariethoz, Felix Liechti, and Baptiste Schmid.
\newblock Quantifying year-round nocturnal bird migration with a fluid dynamics model.
\newblock {\em Journal of the Royal Society Interface}, 18(179):20210194, 2021.

\bibitem{cooper2023songbirds}
Nathan~W Cooper, Bryant~C Dossman, Lucas~E Berrigan, J~Morgan Brown, Alicia~R Brunner, Helen~E Chmura, Dominic~A Cormier, Camille B{\'e}gin-Marchand, Amanda~D Rodewald, Philip~D Taylor, et~al.
\newblock Songbirds initiate migratory flights synchronously relative to civil dusk.
\newblock {\em Movement Ecology}, 11(1):24, 2023.

\bibitem{gauthreaux2003using}
Sidney~A Gauthreaux, Carroll~G Belser, and Donald Van~Blaricom.
\newblock Using a network of wsr-88d weather surveillance radars to define patterns of bird migration at large spatial scales.
\newblock In {\em Avian migration}, pages 335--346. Springer, 2003.

\bibitem{gauthreaux2006monitoring}
Sidney~A Gauthreaux~Jr and John~W Livingston.
\newblock Monitoring bird migration with a fixed-beam radar and a thermal-imaging camera.
\newblock {\em Journal of Field Ornithology}, 77(3):319--328, 2006.

\bibitem{farnsworth2016characterization}
Andrew Farnsworth, Benjamin~M Van~Doren, Wesley~M Hochachka, Daniel Sheldon, Kevin Winner, Jed Irvine, Jeffrey Geevarghese, and Steve Kelling.
\newblock A characterization of autumn nocturnal migration detected by weather surveillance radars in the northeastern usa.
\newblock {\em Ecological Applications}, 26(3):752--770, 2016.

\bibitem{van2018continental}
Benjamin~M Van~Doren and Kyle~G Horton.
\newblock A continental system for forecasting bird migration.
\newblock {\em Science}, 361(6407):1115--1118, 2018.

\bibitem{farnsworth2005flight}
Andrew Farnsworth.
\newblock Flight calls and their value for future ornithological studies and conservation research.
\newblock {\em The Auk}, 122(3):733--746, 2005.

\bibitem{efford2009population}
Murray~G Efford, Deanna~K Dawson, and David~L Borchers.
\newblock Population density estimated from locations of individuals on a passive detector array.
\newblock {\em Ecology}, 90(10):2676--2682, 2009.

\bibitem{challeat2024dataset}
Samuel Chall{\'e}at, Nicolas Farrugia, J{\'e}r{\'e}my~SP Froidevaux, Amandine Gasc, and Nicolas Pajusco.
\newblock a dataset of acoustic measurements from soundscapes collected worldwide during the covid-19 pandemic.
\newblock {\em Scientific Data}, 11(1):928, 2024.

\bibitem{kahl2021birdnet}
Stefan Kahl, Connor~M Wood, Maximilian Eibl, and Holger Klinck.
\newblock Birdnet: A deep learning solution for avian diversity monitoring.
\newblock {\em Ecological Informatics}, 61:101236, 2021.

\bibitem{perch}
Google research. google bird vocalization classifier: A global bird embedding and classification model.
\newblock \url{https://tfhub.dev/google/bird-vocalization-classifier/4}.

\bibitem{van2024nighthawk}
Benjamin~M Van~Doren, Andrew Farnsworth, Kate Stone, Dylan~M Osterhaus, Jacob Drucker, and Grant Van~Horn.
\newblock Nighthawk: acoustic monitoring of nocturnal bird migration in the americas.
\newblock {\em Methods in Ecology and Evolution}, 15(2):329--344, 2024.

\bibitem{rauch2024birdset}
Lukas Rauch, Raphael Schwinger, Moritz Wirth, Ren{\'e} Heinrich, Jonas Lange, Stefan Kahl, Bernhard Sick, Sven Tomforde, and Christoph Scholz.
\newblock Birdset: A multi-task benchmark for classification in avian bioacoustics.
\newblock {\em arXiv preprint arXiv:2403.10380}, 2024.

\bibitem{xenoc}
Xeno-canto foundation and naturalis biodiversity center.
\newblock \url{https://xeno-canto.org}.

\bibitem{macaulay}
The cornell lab of ornithology. macaulay library.
\newblock \url{https://www.macaulaylibrary.org}.

\bibitem{michaud2023unsupervised}
F{\'e}lix Michaud, J{\'e}r{\^o}me Sueur, Maxime Le~Cesne, and Sylvain Haupert.
\newblock Unsupervised classification to improve the quality of a bird song recording dataset.
\newblock {\em Ecological Informatics}, 74:101952, 2023.

\bibitem{morfi2019nips4bplus}
Veronica Morfi, Yves Bas, Hanna Pamu{\l}a, Herv{\'e} Glotin, and Dan Stowell.
\newblock Nips4bplus: a richly annotated birdsong audio dataset.
\newblock {\em PeerJ Computer Science}, 5:e223, 2019.

\bibitem{bc_colombia}
Alvaro Vega-Hidalgo, Stefan Kahl, Laurel~B. Symes, Viviana Ruiz-Gutiérrez, Ingrid Molina-Mora, Fernando Cediel, Luis Sandoval, and Holger Klinck.
\newblock A collection of fully-annotated soundscape recordings from neotropical coffee farms in colombia and costa rica.
\newblock \url{https://zenodo.org/records/7525349}.

\bibitem{bc_amazon}
W.~Alexander Hopping, Stefan Kahl, and Holger Klinck.
\newblock A collection of fully-annotated soundscape recordings from the southwestern amazon basin.
\newblock \url{https://zenodo.org/records/7079124}.

\bibitem{bc_ne_us}
Stefan Kahl, Russell Charif, and Holger Klinck.
\newblock A collection of fully-annotated soundscape recordings from the northeastern united states.
\newblock \url{https://zenodo.org/records/7018484}.

\bibitem{bc_hawaii}
Amanda Navine, Stefan Kahl, Ann Tanimoto-Johnson, Holger Klinck, and Patrick Hart.
\newblock A collection of fully-annotated soundscape recordings from the island of hawai'i.
\newblock \url{https://zenodo.org/records/7078499}.

\bibitem{bc_sierra}
Mary Clapp, Stefan Kahl, Erik Meyer, Megan McKenna, Holger Klinck, and Gail Patricelli.
\newblock A collection of fully-annotated soundscape recordings from the southern sierra nevada mountain range.
\newblock \url{https://zenodo.org/records/7525805}.

\bibitem{bc_w_us}
Stefan Kahl, Connor~M. Wood, Philip Chaon, M.~Zachariah~Peery, and Holger Klinck.
\newblock A collection of fully-annotated soundscape recordings from the western united states.
\newblock \url{https://zenodo.org/records/7050014}.

\bibitem{shrestha2021bird}
Roman Shrestha, Cornelius Glackin, Julie Wall, and Nigel Cannings.
\newblock Bird audio diarization with faster r-cnn.
\newblock In {\em International Conference on Artificial Neural Networks}, pages 415--426. Springer, 2021.

\bibitem{binley2025quantifying}
Allison~D Binley, Jeffrey~O Hanson, Orin~J Robinson, Gregory~H Golet, and Joseph~R Bennett.
\newblock Quantifying the value of participatory science data for conservation decision-making.
\newblock {\em Journal of Applied Ecology}, 2025.

\bibitem{ren2016faster}
Shaoqing Ren, Kaiming He, Ross Girshick, and Jian Sun.
\newblock Faster r-cnn: Towards real-time object detection with region proposal networks.
\newblock {\em IEEE transactions on pattern analysis and machine intelligence}, 39(6):1137--1149, 2016.

\bibitem{lin2017feature}
Tsung-Yi Lin, Piotr Doll{\'a}r, Ross Girshick, Kaiming He, Bharath Hariharan, and Serge Belongie.
\newblock Feature pyramid networks for object detection.
\newblock In {\em Proceedings of the IEEE conference on computer vision and pattern recognition}, pages 2117--2125, 2017.

\bibitem{kahl2023overview}
Stefan Kahl, Tom Denton, Holger Klinck, Hendrik Reers, Francis Cherutich, Herv{\'e} Glotin, Herv{\'e} Go{\"e}au, Willem-Pier Vellinga, Robert Planqu{\'e}, and Alexis Joly.
\newblock Overview of birdclef 2023: Automated bird species identification in eastern africa.
\newblock In {\em CLEF (Working Notes)}, pages 1934--1942, 2023.

\bibitem{swaminathan2024multi}
Bhuvaneswari Swaminathan, M~Jagadeesh, and Subramaniyaswamy Vairavasundaram.
\newblock Multi-label classification for acoustic bird species detection using transfer learning approach.
\newblock {\em Ecological Informatics}, 80:102471, 2024.

\bibitem{kahl2022overview}
Stefan Kahl, Amanda Navine, Tom Denton, Holger Klinck, Patrick Hart, Herv{\'e} Glotin, Herv{\'e} Go{\"e}au, Willem-Pier Vellinga, Robert Planqu{\'e}, and Alexis Joly.
\newblock Overview of birdclef 2022: Endangered bird species recognition in soundscape recordings.
\newblock In {\em CLEF (Working Notes)}, pages 1929--1939, 2022.

\bibitem{arriaga2015bird}
Julio~G Arriaga, Martin~L Cody, Edgar~E Vallejo, and Charles~E Taylor.
\newblock Bird-db: A database for annotated bird song sequences.
\newblock {\em Ecological Informatics}, 27:21--25, 2015.

\bibitem{lostanlen2018birdvox}
Vincent Lostanlen, Justin Salamon, Andrew Farnsworth, Steve Kelling, and Juan~Pablo Bello.
\newblock Birdvox-full-night: A dataset and benchmark for avian flight call detection.
\newblock In {\em 2018 IEEE international conference on acoustics, speech and signal processing (ICASSP)}, pages 266--270. IEEE, 2018.

\bibitem{gomez2023western}
Joan G{\'o}mez-G{\'o}mez, Ester Vida{\~n}a-Vila, and Xavier Sevillano.
\newblock Western mediterranean wetland birds dataset: A new annotated dataset for acoustic bird species classification.
\newblock {\em Ecological Informatics}, 75:102014, 2023.

\bibitem{jamil2023siulmalaya}
Nursuriati Jamil, Ahmad~Nazem Norali, Muhammad~Izzad Ramli, Ahmad Khusaini Mohd~Kharip Shah, and Ismail Mamat.
\newblock Siulmalaya: an annotated bird audio dataset of malaysia lowland forest birds for passive acoustic monitoring.
\newblock {\em Bulletin of Electrical Engineering and Informatics}, 12(4):2269--2281, 2023.

\bibitem{bioac_db}
Datasets for bioacoustics.
\newblock \url{https://bioacoustic-ai.github.io/bioacoustics-datasets/}.

\bibitem{audacity}
Audacity.
\newblock \url{https://www.audacityteam.org/}.

\bibitem{redmon2016you}
Joseph Redmon, Santosh Divvala, Ross Girshick, and Ali Farhadi.
\newblock You only look once: Unified, real-time object detection.
\newblock In {\em Proceedings of the IEEE conference on computer vision and pattern recognition}, pages 779--788, 2016.

\bibitem{lin2017focal}
T~Lin.
\newblock Focal loss for dense object detection.
\newblock {\em arXiv preprint arXiv:1708.02002}, 2017.

\bibitem{tan2020efficientdet}
Mingxing Tan, Ruoming Pang, and Quoc~V Le.
\newblock Efficientdet: Scalable and efficient object detection.
\newblock In {\em Proceedings of the IEEE/CVF conference on computer vision and pattern recognition}, pages 10781--10790, 2020.

\bibitem{carion2020end}
Nicolas Carion, Francisco Massa, Gabriel Synnaeve, Nicolas Usunier, Alexander Kirillov, and Sergey Zagoruyko.
\newblock End-to-end object detection with transformers.
\newblock In {\em European conference on computer vision}, pages 213--229. Springer, 2020.

\bibitem{reis2023real}
Dillon Reis, Jordan Kupec, Jacqueline Hong, and Ahmad Daoudi.
\newblock Real-time flying object detection with yolov8.
\newblock {\em arXiv preprint arXiv:2305.09972}, 2023.

\bibitem{girshick2015region}
Ross Girshick, Jeff Donahue, Trevor Darrell, and Jitendra Malik.
\newblock Region-based convolutional networks for accurate object detection and segmentation.
\newblock {\em IEEE transactions on pattern analysis and machine intelligence}, 38(1):142--158, 2015.

\bibitem{girshick2015fast}
R~Girshick.
\newblock Fast r-cnn.
\newblock {\em arXiv preprint arXiv:1504.08083}, 2015.

\bibitem{zhu2020deformable}
Xizhou Zhu, Weijie Su, Lewei Lu, Bin Li, Xiaogang Wang, and Jifeng Dai.
\newblock Deformable detr: Deformable transformers for end-to-end object detection.
\newblock {\em arXiv preprint arXiv:2010.04159}, 2020.

\bibitem{tan2021efficientnetv2}
Mingxing Tan and Quoc Le.
\newblock Efficientnetv2: Smaller models and faster training.
\newblock In {\em International conference on machine learning}, pages 10096--10106. PMLR, 2021.

\bibitem{liu2018path}
Shu Liu, Lu~Qi, Haifang Qin, Jianping Shi, and Jiaya Jia.
\newblock Path aggregation network for instance segmentation.
\newblock In {\em Proceedings of the IEEE conference on computer vision and pattern recognition}, pages 8759--8768, 2018.

\bibitem{bahdanau2014neural}
Dzmitry Bahdanau.
\newblock Neural machine translation by jointly learning to align and translate.
\newblock {\em arXiv preprint arXiv:1409.0473}, 2014.

\bibitem{vaswani2017attention}
A~Vaswani.
\newblock Attention is all you need.
\newblock {\em Advances in Neural Information Processing Systems}, 2017.

\bibitem{tan2020efficientnetrethinkingmodelscaling}
Mingxing Tan and Quoc~V. Le.
\newblock Efficientnet: Rethinking model scaling for convolutional neural networks, 2020.

\end{thebibliography}

\end{document}